\documentclass[aps, pre, preprint, superscriptaddress, showpacs, showkeys]{revtex4-1}
\usepackage{amsthm}
\usepackage{subfigure}
\usepackage{graphicx}
\usepackage{epsfig}
\usepackage{color}
\usepackage{caption}
\usepackage{amsmath}
\usepackage{amsthm}
\usepackage{amssymb}
\usepackage{hyperref}
\usepackage{ulem}
\newcommand\solidrule[1][1cm]{\rule[0.5ex]{#1}{.4pt}}
\newcommand\dashedrule{\mbox{%
  \solidrule[2mm]\hspace{2mm}\solidrule[2mm]\hspace{2mm}\solidrule[2mm]}}
\newcommand\dasheddotrule{\mbox{%
  \solidrule[2mm]\hspace{1mm}\textperiodcentered\hspace{1mm}\solidrule[2mm]\hspace{1mm}\textperiodcentered\hspace{1mm}\solidrule[2mm]}}

\begin{document}

\title{Viscosity scaling of fingering instability in finite slices with Korteweg stress}

\author{Satyajit Pramanik, Manoranjan Mishra}
\affiliation{Department of Mathematics, Indian Institute of Technology Ropar, 140001 Rupnagar, India}


\begin{abstract}
We perform linear stability analyses (LSA) and direct numerical simulations (DNS) to investigate the influence of the dynamic viscosity on viscous fingering (VF) instability in miscible slices. Selecting the characteristic scales appropriately the importance of the magnitude of the dynamic viscosity of individual fluids on VF in miscible slice has been shown in the context of the transient interfacial tension. Further, we have confirmed this result for immiscible fluids and manifest the similarities between VF in immiscible and miscible slices with transient interfacial tension. In a more general setting, the findings of this pletter will be very useful for multiphase viscous flow, in which the momentum balance equation contains an additional stress term free from the dynamic viscosity. 
\end{abstract}

\maketitle

Displacement processes through porous rocks and mixing of two miscible fluids are active areas of research, having several industrial and environmental applications, such as enhanced oil recovery, hydrology and filtration, carbon capture and storage, etc. VF, a hydrodynamic instability that occurs in both the immiscible and miscible fluids while displacing a more viscous fluid by a less viscous one, is inherent in such flow configuration \cite{Homsy, BDR, GS}. In immiscible fluids surface tension force at the interface acts against the instability \cite{ST}. On the other hand, in miscible fluids, where a thermodynamically stable interface does not exist, a transition zone relaxes with time due to diffusion and acts against the finger growth. Experiments \cite{PWTLMPZ, PBVP} reveal, when the diffusion is slow, a steep gradient in the form of density, concentration or temperature between the underlying fluids gives rise to a weak transient interfacial tension that mimic surface tension effect. This was first discussed by Korteweg in 1901 \cite{K}, who introduced an additional stress term, known as the Korteweg stress, in the equation of motion. The existence of Korteweg stress or transient surface tension is also observed in the experiments of colloidal suspensions \cite{TMDC} and in the binary liquid system of isobuteric acid and water \cite{LGBDPK}. 

Chen and Wang \cite{CW} analyzed the influence of VF instability on the spreading of a localized fluid slice having higher mobility than the surrounding fluid. On the other hand, De Wit {\it et al.} \cite{DBM} studied the same problem in the context of separation in a chromatographic column when the viscosity of the sample is higher than the solvent. Mishra {\it et al.} \cite{MMD} have shown that the onset of VF instability and the subsequent finger pattern near the onset are identical for both the less and more viscous slice. Influence of the Korteweg stresses on VF instability was investigated theoretically by Joseph and his co-workers \cite{J, HJ} and Chen {\it et al.} \cite{CWM} However, to the best of the authors' knowledge the influence of such stresses on the nonlinear VF instability of more and less viscous miscible slices in a Hele-Shaw cell has never been addressed adequately. In particular this letter addresses the question, what is the influence of the Korteweg stress that describes volume forces arising because of the nonlocal molecular interactions on the VF at the rear and frontal interfaces of a localized slice? Such a classical complex pattern dynamics has been investigated through a highly accurate Fourier-spectral method based direct numerical simulations \cite{DBM}. It has been proved theoretically that an appropriate choice of the dynamic viscosity of underlying fluids results the identical onset of fingering instability and the subsequent finger patterns are also identical for both more and less viscous slices in the presence of the Korteweg stresses. The DNS results are found to be in excellent agreement with the corresponding LSA. Also, the similarities between the immiscible slices and miscible one with transient interfacial tension have been proved through LSA, which affirms the classical nature of this study. 

\begin{figure}
\centering
\includegraphics[width=5in, keepaspectratio=true, angle=0]{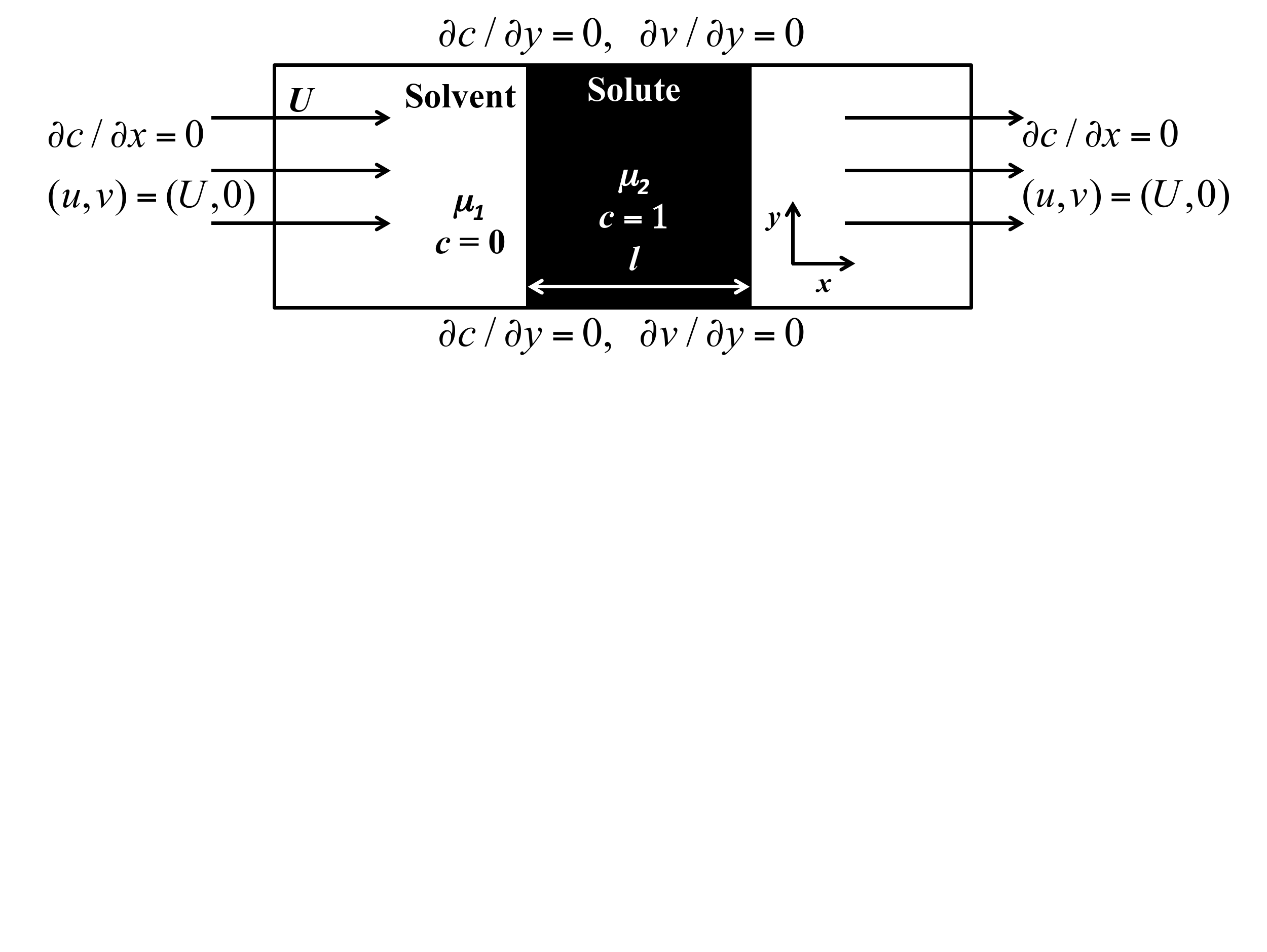}
\vspace{-2.5in}
\caption{Schematic of the flow in a Hele-Shaw cell.}\label{fig:schematic}
\end{figure}

\section{Mathematical formulation and nonlinear simulations}\label{sec:mathformsol}
Consider a uniform rectilinear displacement of a finite fluid slice of viscosity $\mu_2$ by another fluid of viscosity $\mu_1$ in 2D porous media or a Hele-Shaw cell as shown in fig. \ref{fig:schematic}. The frontal interface becomes unstable if $\mu_1 < \mu_2$, otherwise it is the rear interface which features the fingering instability, for $\mu_1 > \mu_2$. The viscosity of the fluids depends on the solute concentration $c$, i.e. $\mu = \mu(c)$. Fluids are assumed to be incompressible and neutrally buoyant. With an additional condition of slow diffusion, the above-mentioned flow problem can be described in terms of Darcy-Korteweg equations \cite{CWM, JR} coupled with a convection-diffusion equation for the mass conservation of the solute concentration. For the dimensionless formulation of the equations the diffusive length and time scales, $D/U$ and $D/U^2$, are used as the respective characteristic scales. Characteristic pressure, velocity, concentration and viscosity are taken to be $\mu_1D/\kappa, U, c_2$ and $\mu_1$, respectively. Here $\kappa$ is the constant permeability of the homogeneous porous medium. For simplicity we have assumed a constant isotropic diffusion of the solute concentration, characterized by the diffusion coefficient $D$. The dimensionless equations in a Lagrangian frame of reference, moving with the speed $U$ are written as,
\begin{eqnarray}
\label{eq:cont}
& & \nabla \cdot \vec{u} = 0, \nabla p = -\mu(c)(\vec{u} + \hat{i}) + \delta\;\nabla\cdot(\nabla c \otimes \nabla c), \\
\label{eq:diff}
& & c_t + \vec{u}\cdot\nabla c = \nabla^2c.
\end{eqnarray}
Here $\vec{u}$ is the gap-averaged velocity having longitudinal and transverse components $u$ and $v$, respectively, $p$ is the dynamic pressure, $\delta  = \hat{\delta}\kappa U^2c_2^2/\mu_1D^3(<0)$ is the dimensionless Korteweg stress constant \cite{CWM, HJ} and the operator $\nabla \equiv \hat{i}\frac{\partial}{\partial x} + \hat{j}\frac{\partial}{\partial y}$. The governing equations (\ref{eq:cont}) - (\ref{eq:diff}) are associated with the following boundary conditions: at the longitudinal boundaries, $\vec{u} = (0,0), \partial c/\partial x = 0, x \to \pm \infty$; and at the transverse boundaries, $\partial v/\partial y = 0$ (representing the constant pressure), $\partial c/\partial y = 0, \forall x$, in the Lagrangian frame of reference (shown in fig. \ref{fig:schematic}). The initial velocity is considered to be $\vec{u} = (0,0)$, while the initial distribution of the solute concentration is, $c = 1$ inside the finite slice and $c = 0$ outside of that. The relationship between the dynamic viscosity of the underlying fluids and the solute concentration is assumed to be of Arrhenius type, $\mu(c) = e^{Rc},$ where $R = \ln (\mu_2/\mu_1)$ is the log-mobility ratio \cite{DBM, Homsy}. Hence, the displacement of a less (more) viscous slice by a more (less) viscous ambient fluid is represented by $R < 0$ ($R > 0$). 

Direct numerical simulations of eqs. (\ref{eq:cont})-(\ref{eq:diff}) in terms of the stream function, $\psi(x,y,t)$ ($u = \psi_y, v = -\psi_x$), 
\begin{eqnarray}
\label{eq:VS1}
& & \nabla^2\psi = -R\nabla c \cdot (\nabla \psi + \hat{j}) + \frac{\delta}{\mu(c)}\Big[\nabla c \times \nabla\left(\nabla^2 c\right)\Big]\cdot\hat{k}, \nonumber \\
& & \\
\label{eq:VS2}
& & c_t + \psi_yc_x - \psi_xc_y = \nabla^2 c,
\end{eqnarray}
are performed using a highly accurate pseudo-spectral method \cite{DBM}. Here $\hat{k}$ represents a unit vector normal to the $xy$-plane. Numerical simulations are performed in a rectangular domain of dimensionless width $L_y = 1024$, and length $A\cdot L_y = 8192$, where $A = L/H$, the ratio of the length to the width of the domain, is the aspect ratio. A convergence analysis of the numerical method and grid independence have been performed. Nonlinear simulations are carried out in the above-mentioned computational domain with $2048 \times 128$ and $2048 \times 256$ spectra, and it has been observed that the maximum relative error between these two sets of simulations is of $\textit{O}(10^{-3})$. Further refinement of the spatial disretization points in either direction does not alter the dynamics of the fingering pattern. Hence, for optimal computational cost we discretize our computation domain of size $[0, 8192] \times [0, 1024]$ using $2048 \times 128$ spectral points. 

\section{Results and Discussion}\label{sec:RD}
\begin{figure}
\centering
(a) \hspace{3in} (b) \\
\includegraphics[width=3.2in, keepaspectratio=true, angle=0]{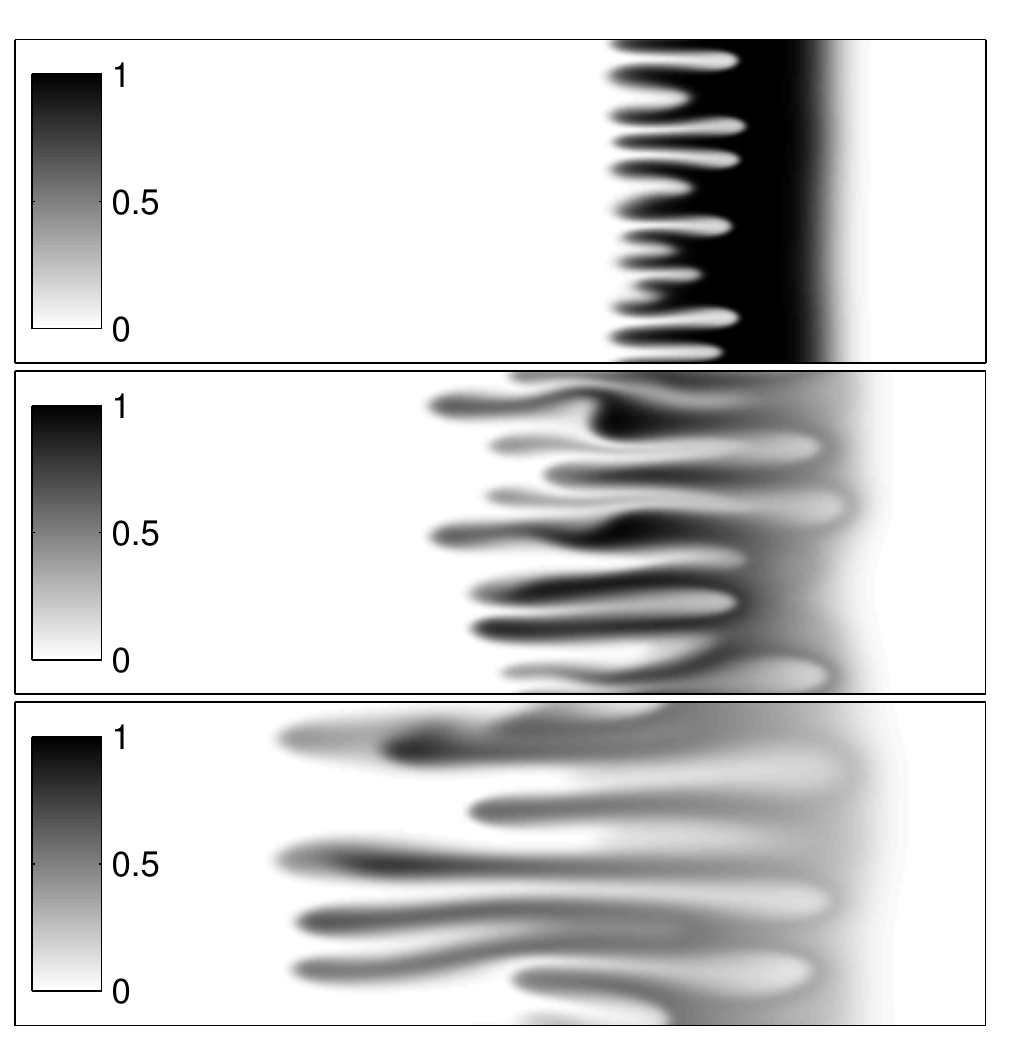}
\includegraphics[width=3.2in, keepaspectratio=true, angle=0]{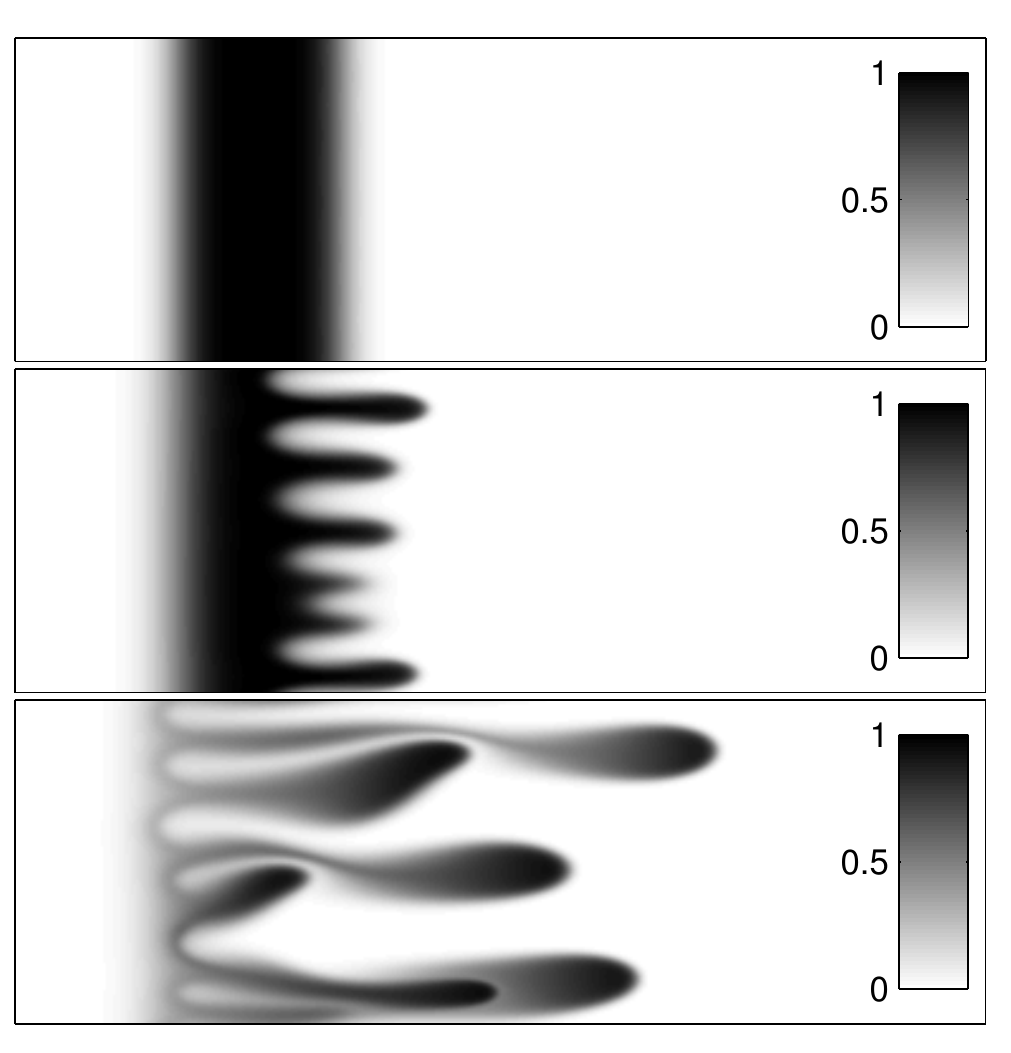} 
\caption{Density plots of concentration at successive times in the Lagrangian reference frame for  $L_y = 1024, l = 256, \delta = -10^4$ when (a) $\mu = e^{3c}$, (b) $\mu = e^{-3c}$. From top to bottom $t = 500, 1000, 1500$. }\label{fig:nonlinear1}
\end{figure}

In the absence of the Korteweg stress VF dynamics for both $R > 0$ and $R < 0$ are the same until the interaction between the two interfaces \cite{MMD}. Our aim in this letter is to determine the possible changes in the dynamics of such slices in the presence of the Korteweg stress and relate it to the surface tension effect in immiscible slices. DNS results for $|R| = 3$, $L_y = 1024, l = 256, \delta = -10^4$ are shown in fig. \ref{fig:nonlinear1}, which depicts that at $t = 500$ fingers appear at the rear interface of the slice for $R = 3$ (see fig. \ref{fig:nonlinear1}(a)), whereas, for $R = -3$ the dynamics of both the interfaces feature diffusive characteristic, but no finger formation at the frontal interface having the unstable viscosity gradient (fig. \ref{fig:nonlinear1}(b)). This signifies that the presence of the Korteweg stresses induces different onset of finger formation for $R > 0$ and $R < 0$. Moreover, the wavelength of the unstable modes is larger for $R < 0$ than that corresponding to $R > 0$. These differences between $R > 0$ and $R < 0$ persist in the long-time behavior of the fingering dynamics of the slices. 

It is customary to ask what causes such differences in the dynamics in the presence of the Korteweg stress? Conventionally, the $\mu-c$ relation, $\mu = e^{Rc}$, keeps $\mu_2/\mu_1$ unchanged for both more and less viscous finite slices, but the dimensionless value of the less (more) viscous fluid changes with the sign of $R$. We notice that in the absence of transient interfacial tension eq. (\ref{eq:VS1}) is free from $\mu(c)$ (see eq. (9) in \cite{MMD}). Hence, the onset of fingering and the qualitative as well as the quantitative pattern of the fingers are identical for $R > 0$ and $R < 0$ in the absence of the Korteweg stress. 

\begin{figure}
\centering
(a) \hspace{3in} (b)
\includegraphics[width=3.2in, keepaspectratio=true, angle=0]{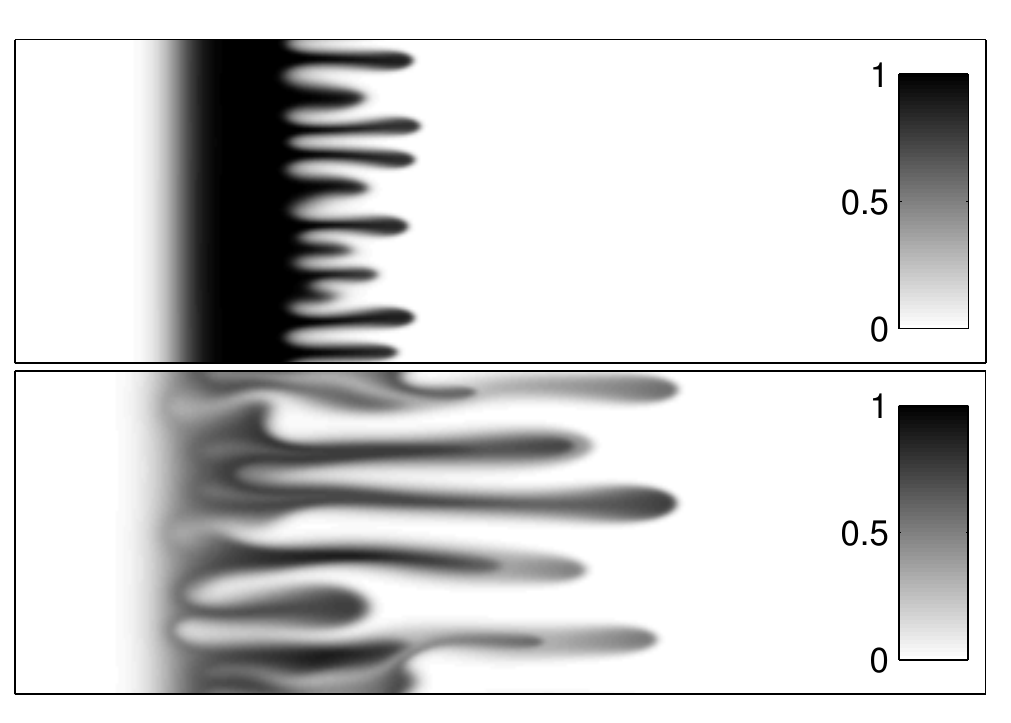}
\includegraphics[width=3.2in, keepaspectratio=true, angle=0]{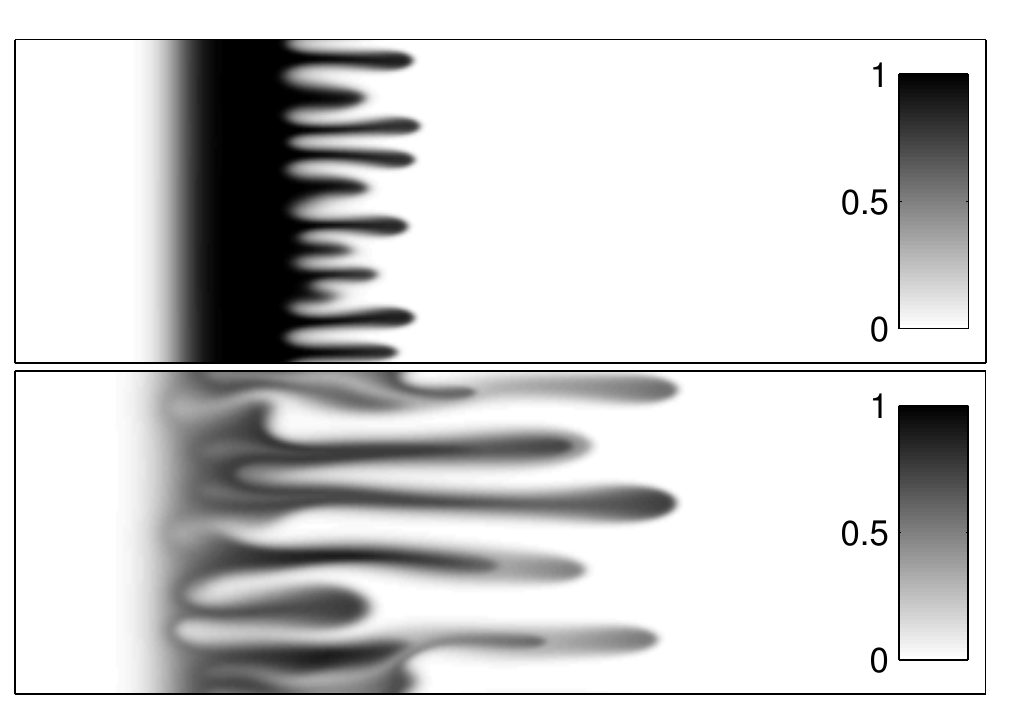}
\caption{Density plots of concentration at successive times, $t = 500$ (top), $t = 1000$ (bottom), in the Lagrangian reference frame for $L_y = 1024, l = 256$: (a) $\delta = -5 \times 10^2, \mu = e^{-3c}$, (b) $\delta = -10^4, \mu = e^{3(1-c)}$. }\label{fig:nonlinear2}
\end{figure}

Choosing the displacing fluid viscosity, $\mu_1$, as the characteristic scale for the dynamic viscosity in both the cases of more and less viscous slices we should be more careful about the dimensionless Korteweg stress constant, $\delta$, which is inversely proportional to the characteristic viscosity. As an example, consider the displacement of a glycerin-water mixture by pure water. The dynamic viscosity of glycerin-water mixture with $68\%$ glycerin concentration is  19.4 cP ($\approx 20$ cP) and that of pure water is 1.002 cP ($\approx 1$ cP). Therefore, in this case we have $\mu_1 \approx1$ cP and $\mu_2 \approx 20$ cP, which can be written in terms of the dimensionless parameter $R = 3$. For this pair of miscible viscous fluids, a dimensional value $\hat{\delta} = -10^{-10}$ N corresponds to the dimensionless value $\delta = -10^4$ for an injection speed $U \approx 10^{-3}$ ms$^{-1}$ and diffusivity $D = 10^{-9}$ m$^2$s$^{-1}$. On the other hand, for the displacement of water by glycerin-water mixture ($C_g = 68\%$) $\mu_1 \approx 20$ cP and $\mu_2 \approx 1$ cP. This makes the dimensionless value $\delta = -5 \times 10^2$ for the same dimensional value $\hat{\delta} = -10^{-10}$ N with the same characteristic variables. In order to justify our claim we perform numerical simulations for $R = -3, \delta = -5 \times 10^2$, and the results obtained are presented graphically in fig. \ref{fig:nonlinear2}(a), which depicts that the onset of instability and the finger pattern, until the interaction of the unstable and the stable interfaces, are identical to those for $R = 3, \delta = -10^4$ (fig. \ref{fig:nonlinear1}(a)). This suggests, in order to compare the numerical results of two similar flow configurations one must choose the dimensionless parameters in such a way that they correspond to the same dimensional value. Subsequently, we ask the question whether there exists a suitable characteristic scale for the dynamic viscosity which automatically takes care of this fact. In particular we choose the less viscosity $\mu_l$ as the characteristic viscosity and $\mu_lD/\kappa$ as the characteristic pressure (see table \ref{tab:viscosity}). Hence, the dimensionless form of $\mu-c$ relation becomes,
\begin{eqnarray}
\label{eq:visco_new1}
& & \mu(c) = e^{Rf(c)}, 
~~~
f(c) = \left\lbrace
\begin{array}{l l}
c, & \mu_1 < \mu_2 \\
1 - c, & \mu_1 > \mu_2, \\
\end{array}
\right.
\\
\label{eq:visco_new2}
& & R = \left\{
\begin{array}{l l}
\ln\left(\displaystyle\mu_2/\mu_1\right), & \mu_1 < \mu_2 \\
\ln\left(\displaystyle\mu_1/\mu_2\right), & \mu_1 > \mu_2.
\end{array}
\right.
\end{eqnarray}
With this modifications we have $R > 0$ for both the more and less viscous slices, and the initial dimensionless dynamic viscosity of the ambient fluid (finite slice) becomes $1$ and $e^R$ ($e^R$ and $1$), respectively (see table \ref{tab:viscosity}). Thus the magnitude of the characteristic viscosity, $\mu_{\mbox{char}}$, remains the same for both the less and more viscous slices that results the same $\delta$ corresponding to every single dimensional value $\hat{\delta}$, in both the cases. Since, the mathematical description of the displacement of a high viscous slice by a low viscous ambient fluid remains the same after the modification, in what follows is the discussion of the displacement of a low viscous slice by a high viscous ambient fluid to understand the consequences of such modifications on the VF instability of more and less viscous slices in the presence of the Korteweg stresses. Fig. \ref{fig:nonlinear2}(b) shows the concentration distributions at successive times for $\mu = e^{3(1 - c)}$, $L_y =1024, l = 256, \delta = -10^4$, which are identical to those in fig. \ref{fig:nonlinear2}(a), hence depicts that the onset of VF and finger pattern are the same to those for a more viscous slice (fig. \ref{fig:nonlinear1}(a)), until the interaction of the fingers with the respective stable interface. 

\begin{table}[!h]
\label{tab:viscosity}
\begin{center}
\begin{tabular}{lccccc}
  \hline
      $\mu_1$(cP) & $\mu_2$(cP) & $\mu_{\mbox{char}}$ & $\mu_1^d$ & $\mu_2^d$ & $R$\\ [3pt]
      \hline
       1.002 & 19.4 & $\mu_1$ & 1 & 19.3613 & $\approx$ 3\\
       19.4 & 1.002 & $\mu_1$ & 1/19.3613 & 1 & $\approx$ -3\\
       19.4 & 1.002 & $\mu_2$ & 19.3613 & 1 & $\approx$ 3\\
  \hline
\end{tabular}
\end{center}
\caption{Dimensional and dimensionless (superscripted `$d$') values of the dynamics viscosity of the underlying fluids, characteristic viscosity, $\mu_{\mbox{char}}$, and $R$ used in the computations. }
\end{table}

\begin{figure}[!h]
\centering
\includegraphics[width=5in, keepaspectratio=true, angle=0]{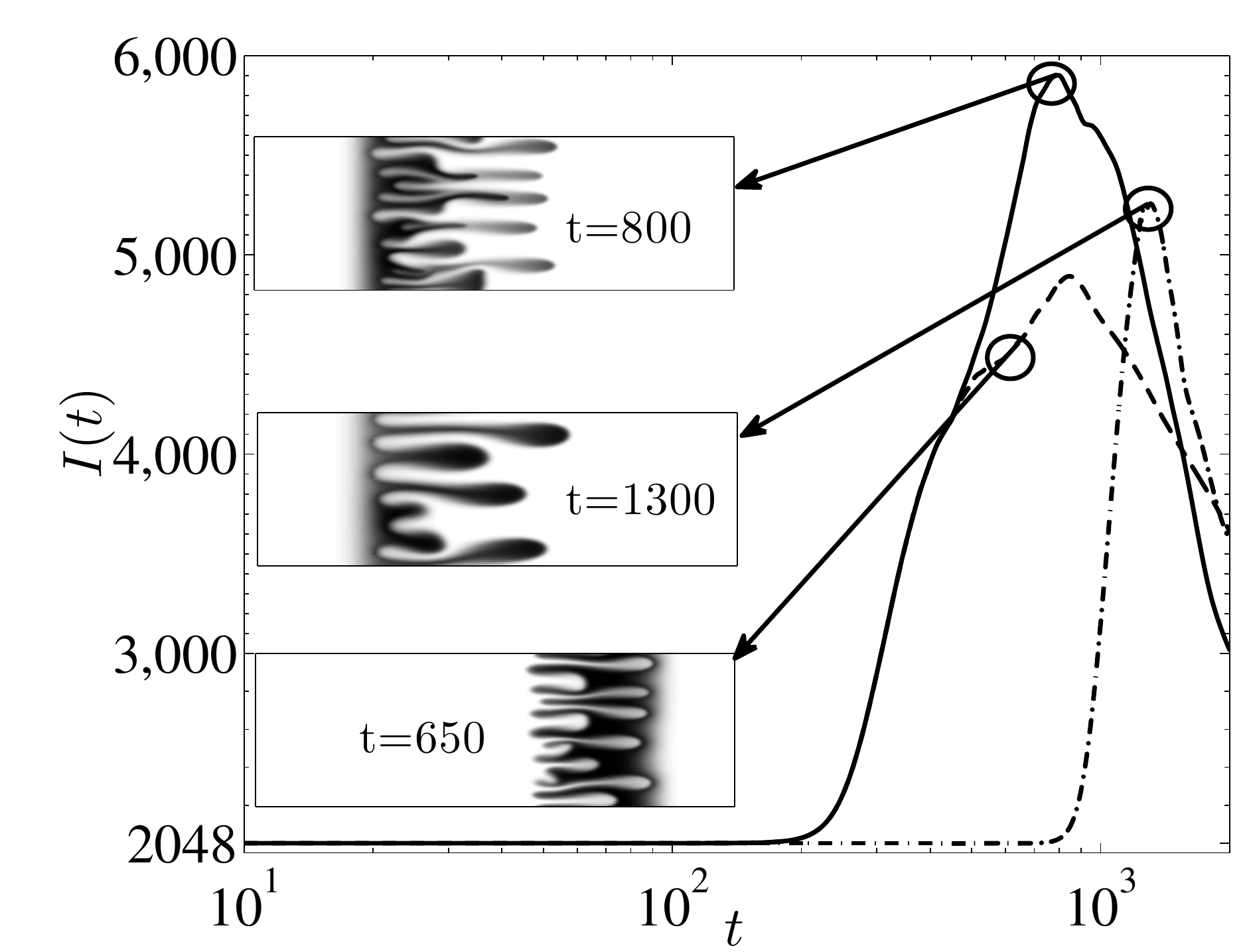}\caption{Interfacial length $I(t)$ for $\mu = e^{3c}$, ($\solidrule$) $\mu = e^{-3c}$ ($\dasheddotrule$) and $\mu = e^{3(1 - c)}$ ($\dashedrule$). The inset images show the concentration distribution approximately at time of interactions between the stable and unstable interfaces.}\label{fig:nonlinear_IL}
\end{figure}

The onset of fingering instability and the interaction between the stable and unstable interfaces can be quantified appropriately from the temporal evolution of the interfacial length \cite{CM, MMD}, $I(t) = \int_0^{L_y}\int_0^{A\cdot L_y} \left(\frac{\partial c}{\partial x}\right)^2 + \left(\frac{\partial c}{\partial y}\right)^2 \mbox{d}x\mbox{d}y$. In the diffusion dominated regime $I(t)$ retains a constant value equal to the width of the domain, and increases with the growth of the fingers. Hence, the onset of fingering is marked as the instance when $I(t)$ starts increasing from the constant value of the diffusive regime. Fig. \ref{fig:nonlinear_IL} shows the temporal evolution of the interfacial length corresponding to the simulations of figs. \ref{fig:nonlinear1} and \ref{fig:nonlinear2}. It depicts that the displacements remain stable over the first two decades of time and the fingers form at $t \approx 180$, for the viscosity relations, $\mu(c) = e^{3c}$ and $\mu(c) = e^{3(1-c)}$, while at $t \approx 800$ for $\mu(c) = e^{-3c}$. $I(t)$ decreases with time after the interaction of the fingers with the respective stable interface. Inset images show the concentration distributions approximately at such instance of interaction and depict how this time differ from each other for different viscosity profiles. Hence, all the qualitative features of the dynamics observed in the absence of any gradient stress \cite{MMD}, can be seen even with the gradient stresses, but with different quantitative measures due to the influence of transient interfacial tension. 

\begin{figure}[!h]
\centering
(a) \hspace{3in} (b) \\
\includegraphics[width=3.2in, keepaspectratio=true, angle=0]{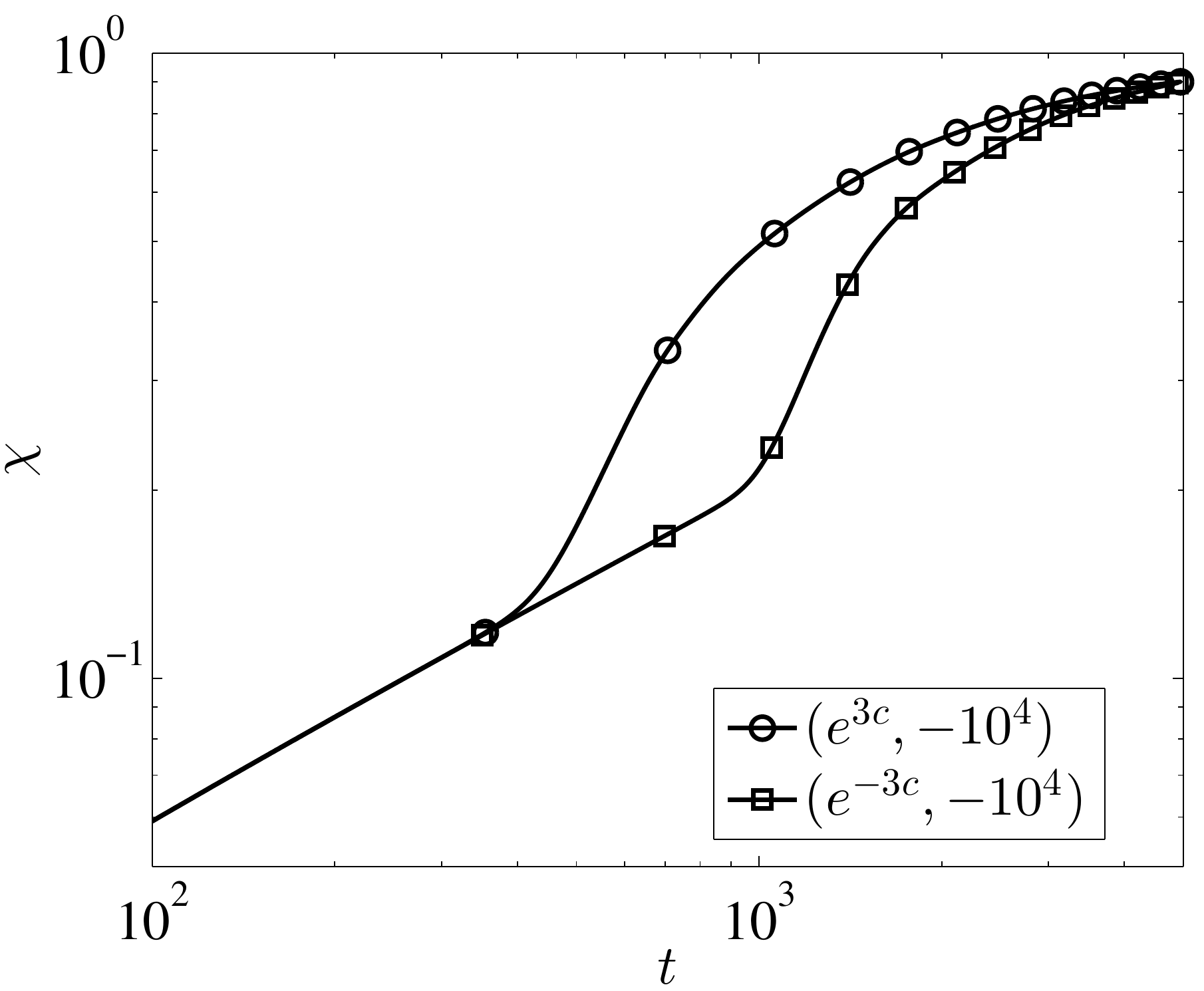}
\includegraphics[width=3.2in, keepaspectratio=true, angle=0]{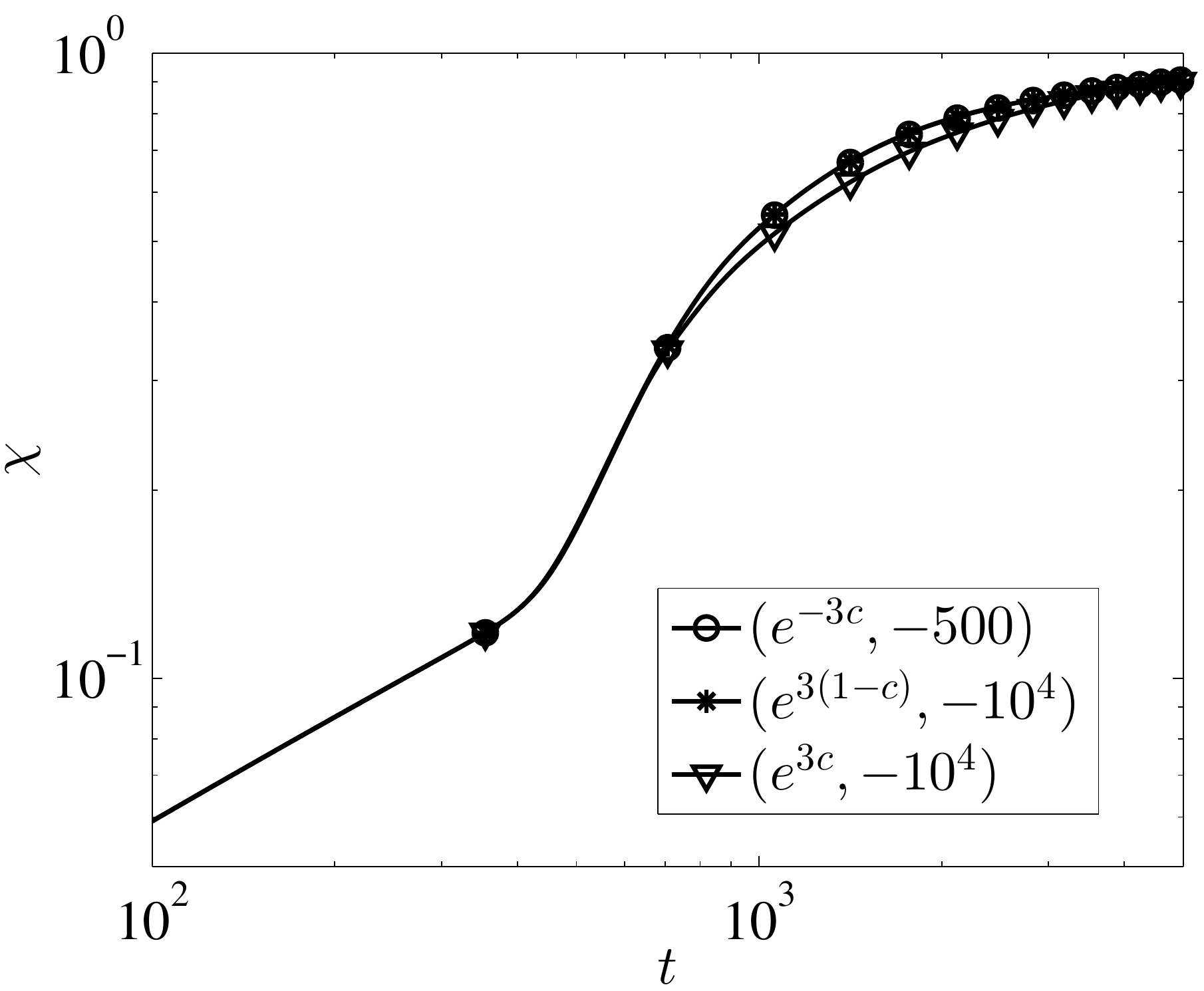} 
\caption{Temporal evolution of $\chi(t)$ for a more and less viscous finite slice of width, $l = 256$ for $L_y = 1024$ and $\delta$. }\label{fig:degree_mixing}
\end{figure}

VF enhances the concentration gradient and the area available for diffusive flux across the interface by stretching the fluid-fluid interface and hence modifies the rate at which mixing occurs. It would be interesting to understand the effect of the Korteweg stresses on the mixing of a more or less viscous finite slice. The degree of mixing is defined as \cite{JCJ2011}, $\chi(t) = 1 - \sigma^2(t)/\sigma^2_{\mbox{max}}$, in terms of the global variance of the concentration field, $\sigma^2 \equiv \langle c^2 \rangle - \langle c \rangle^2$, where $\langle \cdot \rangle$ represents the spatial averaging over the domain. The degree of mixing of more and less viscous finite slices have been plotted in fig. \ref{fig:degree_mixing} corresponding to the simulations of figs. \ref{fig:nonlinear1} - \ref{fig:nonlinear2}. Due to stronger instability for the case of $R = 3, \delta = -10^4$ than $R = -3, \delta = -10^4$, the degree of mixing, $\chi(t)$, is higher in the former than the latter (see fig. \ref{fig:degree_mixing}(a)). In fig. \ref{fig:degree_mixing}(b), $\chi(t)$ has been shown for the three cases: (i) $\mu = e^{3c}, \delta = -10^4$, (ii) $\mu = e^{-3c}, \delta = -5 \times 10^2$ and (iii) $\mu = e^{3(1-c)}, \delta = -10^4$. It depicts identical temporal evolution of $\chi(t)$ for both more and less viscous slices until the interaction of the fingers with the respective stable interface. Afterwards, $\chi(t)$ becomes higher in the latter case than the former, and it asymptotically saturates to the maximum value of $1$, corresponding to the completely mixed state. The coincidence of the three curves in the time interval $t \sim 650$ confirms the results shown in figs. \ref{fig:nonlinear1}(a), \ref{fig:nonlinear2}. 

\section{Linear stability analysis}\label{sec:LSA}
Here we discuss LSA of miscible slice in the presence of the Korteweg stresses with the viscosity relations given by eqs. (\ref{eq:visco_new1})-(\ref{eq:visco_new2}) and compare the results obtained with those of DNS. Finally, the obtained LSA results are compared with a LSA of immiscible slice which confirms that the Korteweg stress and the surface tension have identical effects on the instability. 

\begin{figure}[!h]
\centering
(a) \hspace{3 in} (b) \\
\includegraphics[width=3.2in, keepaspectratio=true, angle=0]{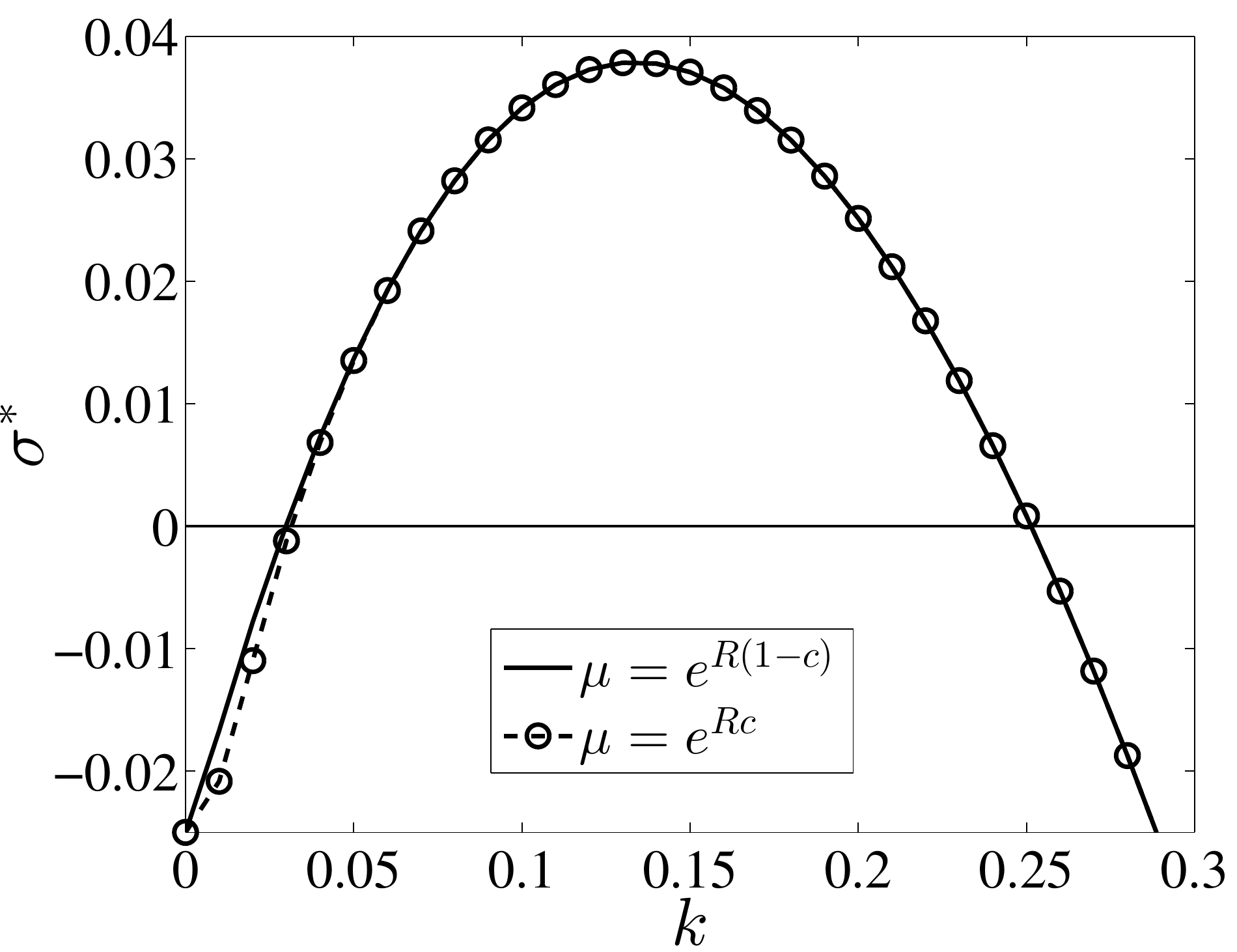}
\includegraphics[width=3.2in, keepaspectratio=true, angle=0]{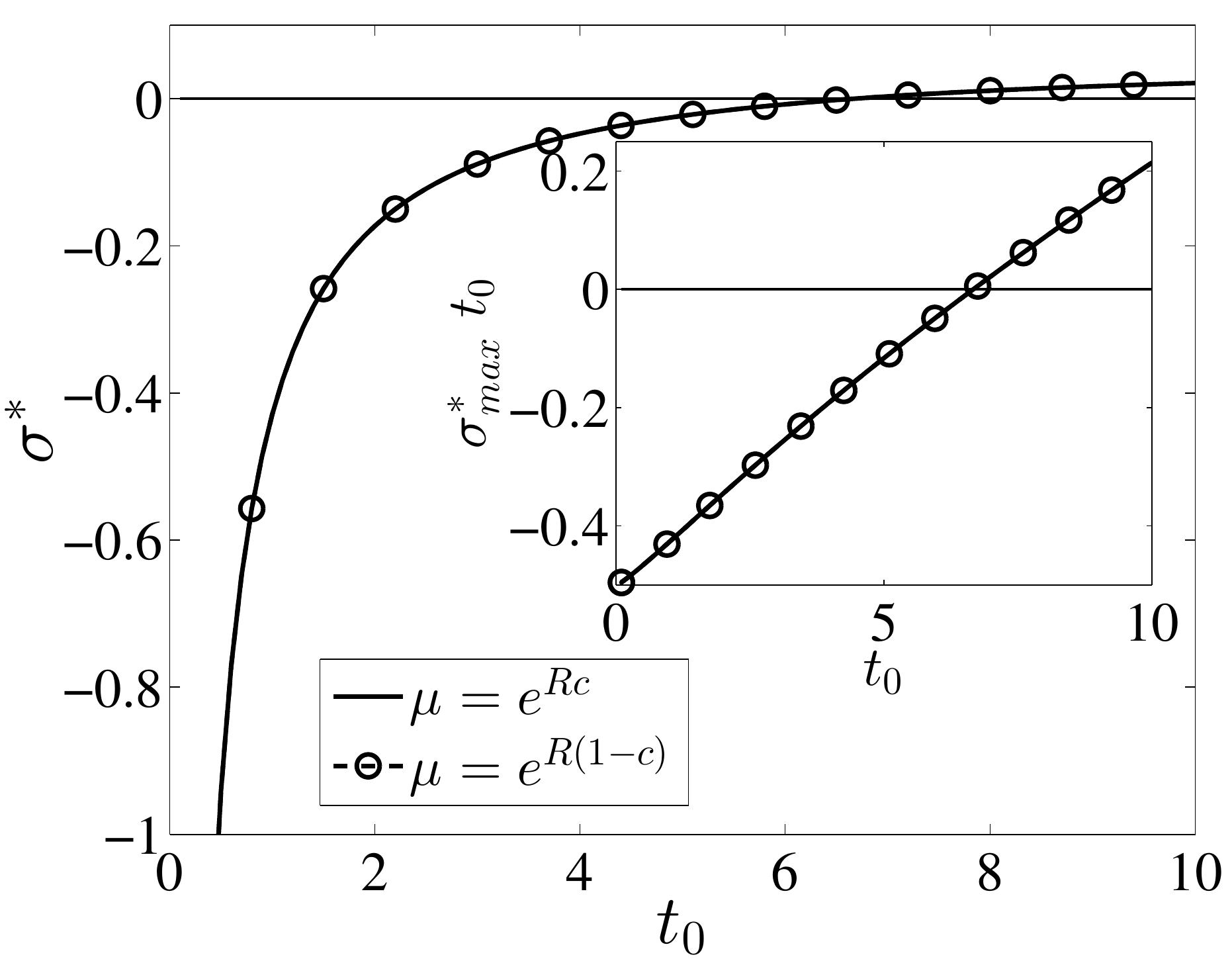}
\caption{(a) Dispersion curves at $t_0 = 20$, (b) Temporal evolution of the growth rate of the most dangerous wave perturbation ($k_m \approx 0.15$). Inset image shows the maximum growth, $\sigma^*_{max}t_0$, from all possible wave number, $k$, representing the onset of instability. The parameter values are $\delta = -10^3, R = 3$ and $l = 100$. }\label{fig:LSA}
\end{figure}

\begin{figure}
\centering
(a) \hspace{3in} (b) \\
\includegraphics[width=2.9in, keepaspectratio=true, angle=0]{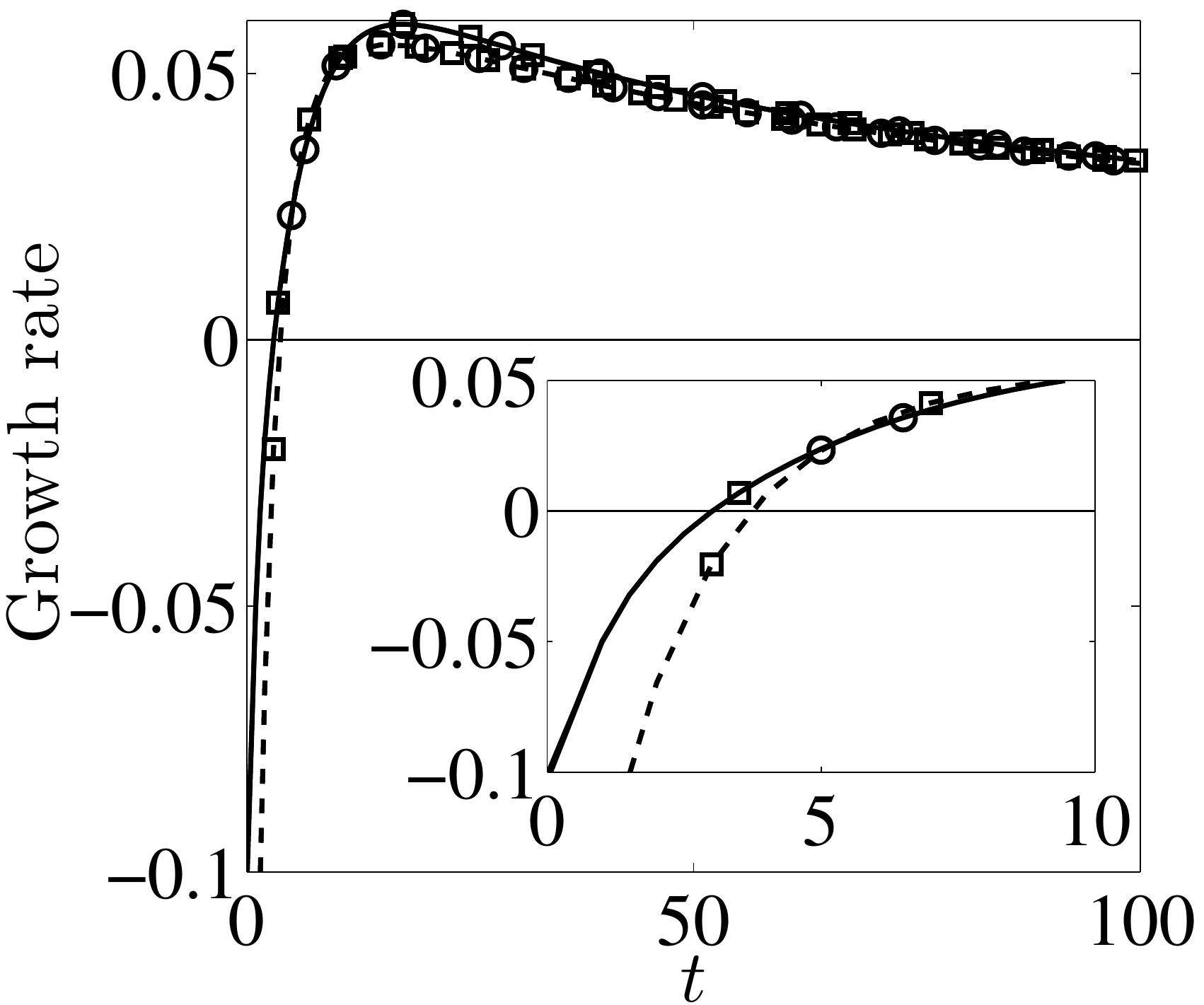}
\includegraphics[width=3.2in, keepaspectratio=true, angle=0]{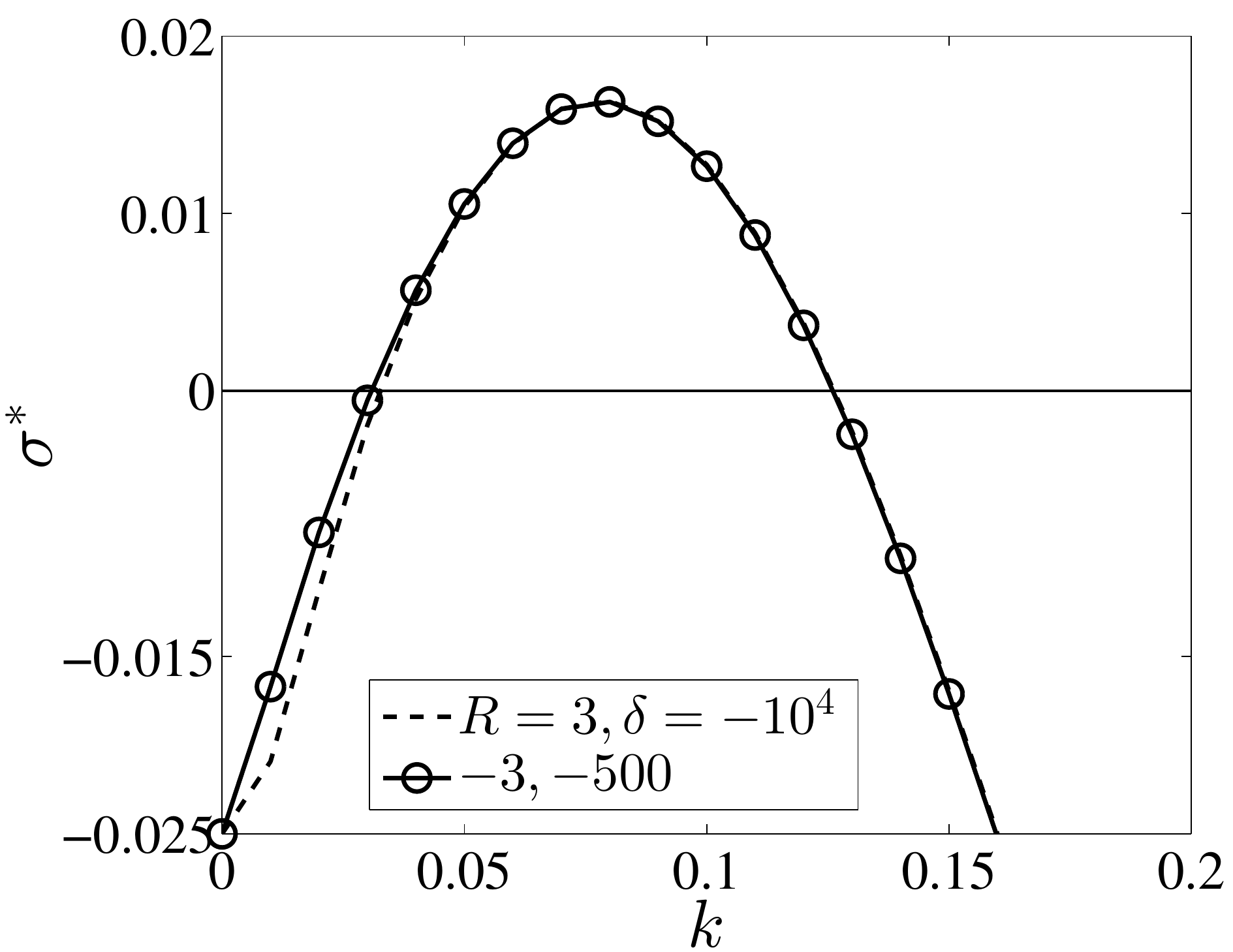} 
\caption{(a) Temporal evolution of the growth rate for $R = 3, l = 100, k = 0.15, \mu = e^{Rc} ~ (\bigcirc), \mu = e^{R(1-c)} ~ (\square), \delta = -10^3$. IVC ($\solidrule$) and DNS ($\dashedrule$). Inset image is magnified near the onset of instability. (b) Dispersion curves at $t_0 = 20$ for $l = 100$ with the self-similar LSA. }\label{fig:comparision}
\end{figure}

\subsection{Miscible slice}\label{subsec:MS}
The initial-boundary value problem described above possesses a self-similar diffusive decaying solution in a similarity transformation $(\xi, t)$-domain \cite{PM}, $\vec{U}_0 = (0,0),~~ \mu_0 = \mu_0(\xi), ~ C_0(\xi) = \big[\mbox{erf}\big(\xi/2\big) - \mbox{erf}\big((\xi - l^*)/2\big)\big]/2$, where $\xi = x/\sqrt{t}$ is the similarity variable. Here, $C_0(\xi)$ corresponds to the self-similar decay of a rectangular function of width $l^* = l/\sqrt{t}$ and height unity. We analyze how an infinitesimal perturbation to this reference solution grows (decays) with time. Eliminating the pressure and transverse velocity component from the linearized form of the eqs. (\ref{eq:cont})-(\ref{eq:diff}), and using the $\mu-c$ relation from eqs. (\ref{eq:visco_new1})-(\ref{eq:visco_new2}), a system of coupled partial differential equations is obtained. These equations are homogeneous in $y$ and $t$, hence we can decompose the perturbation quantities into normal modes as, $(u',c')(\xi,y,t) = (\Phi(\xi),\Psi(\xi))e^{iky + \sigma^*(k,t_0)t}$. Here $\Psi(\xi)$ and $\Phi(\xi)$ are the amplitude of the perturbed concentration $c'$ and axial velocity $u'$, respectively, with $k, \sigma^*(k,t_0)$ being the wave number and the growth rate of the perturbations \cite{PM}. Hence, the linear stability problem can be written as the following eigenvalue problem,
\begin{eqnarray}
\label{eq:coupledode1}
& &\Big(D^2 + f'(c_0) DC_0D-k^{2}t_0 \Big)\Phi = \Big(f'(c_0)k^2t_0 \nonumber \\
& & + (k^{2}\delta/\mu_0\sqrt{t_{0}})\big\{D^3C_0-DC_0\left(D^2 - k^{2}t_0\right)\big\}\Big)\Psi,
\end{eqnarray}
\begin{equation}
\label{eq:coupledode2}
\sqrt{t_{0}}\Big(\sigma^* - (1/t_0)D^2 + k^{2} - (\xi/2t_0)D\Big)\Psi = -DC_0\Phi, 
\end{equation}
where $D^n \equiv \mbox{d}^n/\mbox{d}\xi^n, n \in \mathbb{N}$. 
The system of coupled ordinary differential equations (eqs. (\ref{eq:coupledode1})-(\ref{eq:coupledode2})) has been solved using a finite difference method \cite{PM} to determine the instantaneous growth rate of the perturbations in terms of the wave number $k$ and the frozen diffusive time $t_0$. Dispersion curves obtained from LSA for the displacement of more and less viscous fluid slice are presented in fig. \ref{fig:LSA}(a) for the parameters $R = 3, l = 100, \delta = -10^3$ at the frozen diffusive time $t_0 = 20$. It depicts that the two dispersion curves are almost indistinguishable for the unstable modes. Also, the temporal evolution of the growth rate of the most unstable mode, $k_m= 0.15$ (fig. \ref{fig:LSA}(a)), as well as the onset of instability for all possible modes coincide for $\mu = e^{3c}$ and $\mu = e^{3(1 - c)}$ (fig. \ref{fig:LSA}(b)). Linearizing the stream-function form of eqs. (\ref{eq:cont})-(\ref{eq:diff}) with $\mu-c$ relation given by eqs. (\ref{eq:visco_new1})-(\ref{eq:visco_new2}) and using a pseudo-spectral method, an initial value calculation (IVC) is perfromed. The growth rate of the perturbed concentration, $c'$, $\sigma_{c'} = (1/2E_{c'})(\mbox{d}E_{c'}/\mbox{d}t)$, where $E_{c'} = \int_0^{L_y}\int_0^{A\cdot L_y}c'^2(x,y,t)\mbox{d}x\mbox{d}y$, obtained from IVC is compared with those obtained from DNS \cite{Jessica}. The temporal evolution of the growth rates of IVC coincide with those obtained from DNS for both $\mu = e^{3c}$ and $\mu = e^{3(1-c)}$ with $\delta = -10^3$ (fig. \ref{fig:comparision}(a)). The numerical algorithm of the IVC will be discussed elsewhere in detail. Fig. \ref{fig:comparision}(b) depicts an excellent agreement between the dispersion curves obtained from self-similar LSA at $t_0 = 20$ corresponding to $R = 3, \delta = -10^4$ and $R = -3, \delta = -5 \times 10^2$, and confirms the observation from DNS presented in fig. \ref{fig:nonlinear2}(a). 

\begin{figure}[!h]
\centering
\includegraphics[width=5in, keepaspectratio=true, angle=0]{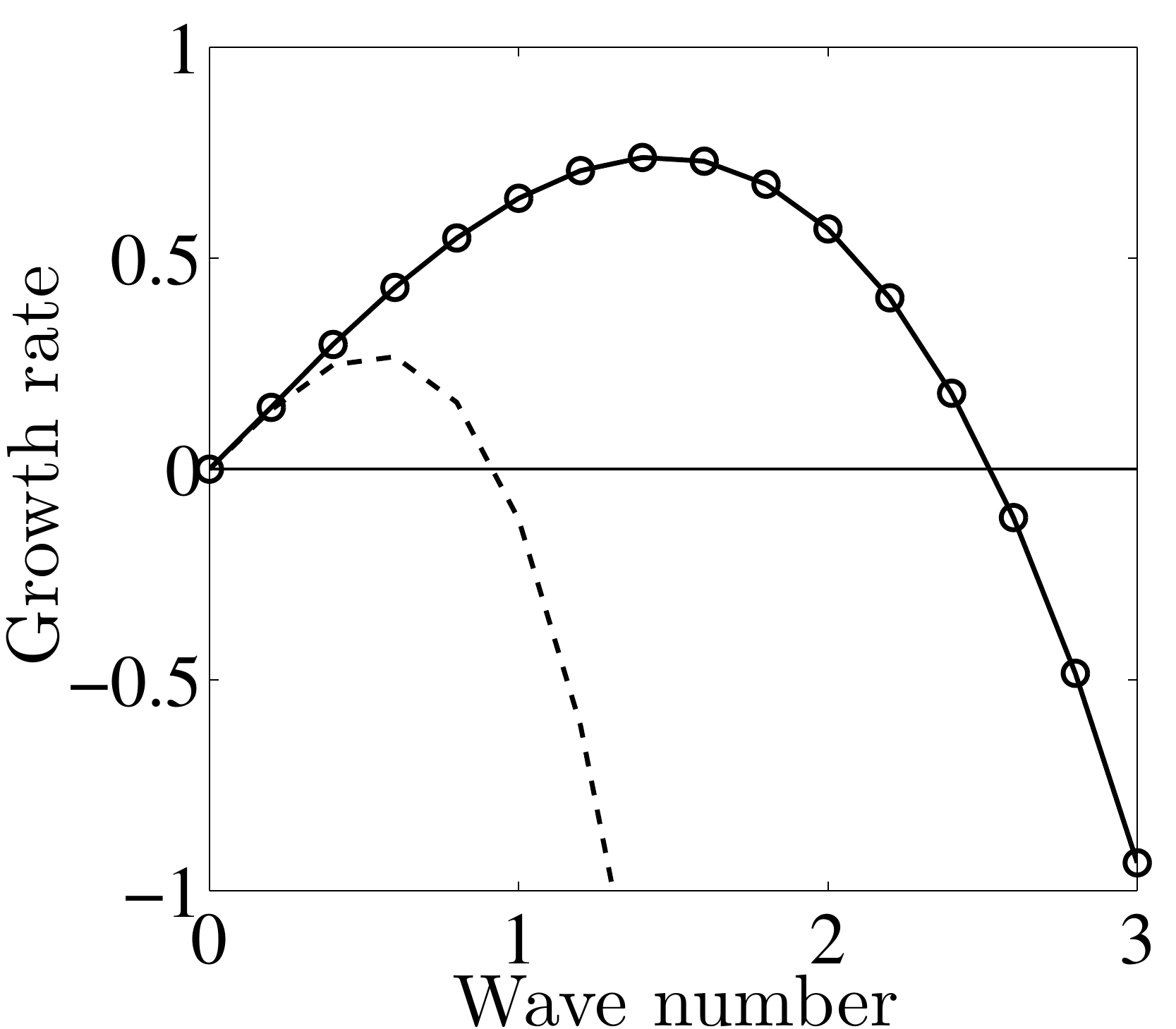}
\caption{Dispersion curves for a three layer immiscible Hele-Shaw flow with $L = 5, S = 1, T = 1, U = 1$ for three sets of viscosities: $\mu_l = \mu_r = e^2, \mu = 1$ ($\solidrule$), $\mu_l = \mu_r = 1, \mu = e^{-2}$ ($\dashedrule$), $\mu_l = \mu_r = 1, \mu = e^{2}$ ($\bigcirc$). }\label{fig:daripaimmiscible}
\end{figure}

\subsection{Immiscible slice}\label{subsec:IS}
In what follows a comparison between an immiscible slice and a miscible slice with the Korteweg stresses. Chen {\it et al.} \cite{CCM} have shown that in a rotating Hele-Shaw cell both the qualitative and quantitative features of the Korteweg stress in miscible fluids and surface tension in immiscible fluids are identical. Not only the stabilizing property of the Korteweg stresses was confirmed, but also the number of fingers were measured to be the same for the miscible and immiscible fluids, both theoretically as well as experimentally. Our aim is to understand the relative importance of the viscous force to the surface tension force in  three layer immiscible displacement, in which the growth rates of the perturbations at the two interfaces can be represented as (see eq. (10) in \cite{Daripa}), $\sigma_{\pm} = (-b \mp \sqrt{b^2 - 4ac})/2a$, where, $a = -e^{kL}(\mu + \mu_l)(\mu + \mu_r) + e^{-kL}(\mu - \mu_l)(\mu - \mu_r) <0, ~ b = [e^{kL}(\mu + \mu_l) + e^{-kL}(\mu - \mu_l)]\xi - [e^{kL}(\mu + \mu_r) + e^{-kL}(\mu - \mu_r)]\tau, ~ c = \tau\xi(e^{kL} - e^{-kL}), ~ \xi = Tk((\mu_r - \mu)U/T - k^2), ~ \tau = Sk(k^2 - (\mu - \mu_l)U/S)$. Here $\sigma_+$ and $\sigma_-$ correspond to the larger and smaller growth rates, respectively. Furthermore, $S$ and $T$ are the surface tension at the left and right interfaces, $L$ being the width of the middle layer, $U$ is the fluid velocity and  $\mu_l, \mu, \mu_r$ are the viscosities of the left, middle and right fluid layers, respectively. These show that the obtained analytic expressions for the dispersion relations at the two interfaces involve explicitly the dynamic viscosities of the three fluid layers, not their ratio. As a consequence, the dispersion curves corresponding to $\sigma_+$ become different for two sets of viscosities although having the same viscosity ratio of two adjacent fluid layers. For instance, we consider three sets of viscosities in such a way that the interface separating the middle layer from the right layer becomes unstable for the first two cases, while for the third, it is the interface between the left and the middle layer that features instability. Corresponding dispersion curves ($\sigma_+$ vs. $k$) are shown in fig. \ref{fig:daripaimmiscible} for $L = 5, S = 1, T = 1, U = 1$; the solid line corresponds to the case of $\mu_l = \mu_r = e^2, \mu = 1$, dashed line corresponding to $\mu_l = \mu_r = 1, \mu = e^{-2}$, and the circles representing the situation when  $\mu_l = \mu_r = 1, \mu = e^{2}$. The viscosities of the left and right layers are chosen to be equal so that it reduces to an analogous form of a finite miscible slice discussed above. Although the ratio of the high viscosity to the low viscosity remain to be the same, $e^2$, the dispersion curve for the second case differs from the remaining two. It signifies that the dispersion curves corresponding to a more and less viscous middle layer will overlap if the dynamic viscosities of the less and more viscous fluids are kept unchanged. Eventually, growth rate is always higher for the case with lager dynamic viscosity of the fluids. Another important observation is the variation of the cutoff wave number ($k_c$) for the unstable mode and the most unstable wave number ($k_m$) with the variation of the surface tension. The qualitative changes of these two quantities are similar to those observed in miscible displacements under the influence of the Korteweg stress. Both in miscible and immiscible fluid systems, $k_c$ and $k_m$ decrease with the increase of the magnitude of $\delta$ and the surface tension force, respectively. 
 
\section{Conclusion}\label{sec:conclu}
We have theoretically investigated the VF instability of miscible slices with transient interfacial tension. An appropriate choice of the dynamic viscosity of the fluid in the characteristic scales manifests its importance in the study of VF with transient interfacial tension. Our analysis is capable of reproducing the results existing in the literature. The results obtained depict that the onset of VF and their dynamical pattern can become identical only under certain scaling analysis for both the more and less viscosity slices in the presence of Korteweg stress, similar to the case when these stresses are absent \cite{MMD}. Synergetic mixing with VF and alternating injection \cite{JCJ2013} will be the same with the present viscosity model irrespective of the choice of the viscosity of the fluid that fills the Hele-Shaw cell before the injection starts. Alike effect of the dynamic viscosity is also observed in immiscible fluids. The findings of this letter will certainly help to understand multiphase viscous flows with different viscosity of each phase or when the viscosity depends non-monotonically on the solute concentration \cite{MH}; for instance, during the mixing process of chemical components, pollutant contamination in aquifers, CO$_2$ sequestration, etc. Miscible displacement of viscous fluids with Korteweg stress has important applications in chemistry \cite{TMDC, LGBDPK}. In this letter we present a preliminary understanding of the relative importance of the Korteweg stress to the viscous stress by incorporating one of the simplest situation of constant diffusivity, $D$, of the solute concentration. A velocity induced or concentration dependent diffusivity should be studied for a more realistic situation. More interesting and challenging problem of miscible VF in the presence of the Korteweg stress induced by both the concentration and temperature gradients is the focus of the authors' future work. 

\acknowledgments
S.P. gratefully thanks the National Board for Higher Mathematics (NBHM), Department of Atomic Energy (DAE), Government of India for the Ph.D. fellowship.

\end{document}